\documentclass[showpacs,amsmath,twocolumn,showkeys,pra,superscriptaddress]{revtex4-1}
\usepackage{dcolumn}    
\usepackage{bm}         
\usepackage{ifpdf}
\usepackage{comment}
\usepackage{amssymb,lineno,amsfonts}
\usepackage{graphicx}   
\usepackage{bbm}
\usepackage{mathrsfs}
\usepackage{upgreek}
\usepackage{epstopdf}
\usepackage{setspace}
\usepackage{hyperref}
\usepackage[matrix,frame,arrow]{xypic}
\usepackage{float}
\usepackage{natbib}
\usepackage{color} 
\usepackage{multirow}
\hypersetup{colorlinks, linkcolor={red},citecolor={blue}, urlcolor={blue}}

\newcommand{\ket}[1]{\vert{#1}\rangle}

\newcommand{\floor}[1]{\left\lfloor #1 \right\rfloor}
\newcommand{\ceil}[1]{\left\lceil #1 \right\rceil}
\newcommand{\f}[1]{C_{#1}}
\begin{document}
\preprint{APS/123-QED}

\title{Hybrid scheme for factorization: \\ Factoring 551 using a 3-qubit NMR quantum adiabatic processor
}
\author{ 
Soham Pal}
\email{soham.pal@students.iiserpune.ac.in}
\affiliation{ 
Department of Physics, Indian Institute of Science Education and Research, Pune, India }
\author{ 
Saranyo Moitra}
\email{saranyo.moitra@gmail.com}
\affiliation{SISSA- International School for Advanced Studies, via Bonomea 265, 34136 Trieste, Italy}
\author{ 
V. S. Anjusha}
\email{anjusha@students.iiserpune.ac.in}
\affiliation{ 
Department of Physics, Indian Institute of Science Education and Research, Pune, India }
\author{ 
Anil Kumar}
\email{anilnmr@iisc.ernet.in}
\affiliation{ 
Department of Physics and NMR Research Centre, Indian Institute of Science, Bangalore-560012}
\author{ 
T. S. Mahesh}
\email{mahesh.ts@iiserpune.ac.in}
\affiliation{ 
Department of Physics, Indian Institute of Science Education and Research, Pune, India }%




\begin{abstract}
Quantum processors are potentially superior to their classical counterparts for many computational tasks including factorization.  Circuit methods as well as adiabatic methods have already been proposed and implemented for finding the factors of a given composite number.    The main challenge in scaling it to larger numbers is the unavailability of large number of qubits.  
Here we propose a hybrid scheme that involves both classical and quantum computation, which reduces the number of qubits required for factorization. 
The classical part involves setting up and partially simplifying a set of bit-wise factoring equations and the quantum part involves solving these coupled equations using a quantum adiabatic process.
We demonstrate the hybrid scheme by factoring 551 using a three qubit NMR quantum register.
	
\end{abstract}

                              \maketitle
                              \author{Soham Pal}


\section{\label{sec:level1}Introduction}

Multiplying two large numbers is an easy task, but the other way, i.e., given a large number, to find its prime factors, is very difficult. In fact there is no known classical algorithm to factor a number with polynomial resources. For many present cryptographic techniques, such as RSA, this fact forms the basis for ensuring secure communication \cite{koblitz}.

Peter Shor in his milestone paper  introduced a quantum algorithm to factorize  numbers with polynomial complexity  \cite{shor1994algorithms,shor1999polynomial}.  Since then, several experimental architectures, including NMR \cite{vandersypen2001experimental}, photonic-systems \cite{photonic_Shor},
and trapped-ions \cite{Monz1068}, have been used to demonstrate Shor's algorithm by factoring small numbers.  
Factoring larger numbers has been hindered by the unavailability of a quantum register with large number of qubits. 
As the size of the quantum register increases, one has to encounter the challenges of increased complexity of qubit-selective quantum controls, decreased coherence times, and  difficulty in quantum measurements.  
More over, it is also believed that the quantum processors may only be as efficient as their classical counter parts in certain computational tasks \cite{nielsen2010quantum}.  In this context, it is practical and may even be advantageous to look for a hybrid processor which can reduce the burden on the quantum processor without compromising the overall efficiency of computation.  

In this work, we provide such an example by describing a hybrid procedure that uses both classical and quantum routines.  We describe factorization of large numbers using two stages: (i) construction and  simplification of bitwise factoring equations using a classical processor, and (ii) solving the bitwise factoring equations using an adiabatic quantum processor.  The adiabatic quantum factorization was previously used to  factor 21 (\cite{pengfactorizationAQC}) and 143 (\cite{xu2012quantum}) using three and four qubits respectively.  The hybrid method allows significant reduction in the number of qubits, and hence the complexity of quantum operations.  Here we describe factorization of 551 using only three qubits.  More over, we experimentally demonstrate the adiabatic solution of bitwise factoring equations using a three-qubit NMR system.  

In the next section we describe the theoretical aspects of the hybrid procedure for factorization.  In section III we describe the NMR experiments to factor 551, and finally we conclude in section IV.

\section{\label{sec:level2}Theory}
Let $n$ be an $l_n$-bit biprime which is to be factored into its two prime factors $p$ and $q$, i.e., $n=p \times q$.  
We can encode the factors on two quantum registers with $l_p$ and $l_q$ qubits.
In binary form, the composite number and its factors are
\begin{equation}
n = \sum_{i=0}^{l_n-1}2^i n_i, ~
p = \sum_{j=0}^{l_p-1}2^j p_j, ~ \mathrm{and} ~
q = \sum_{k=0}^{l_q-1}2^k q_k.
\end{equation}
Except for the cases where one of the factors $p$ or $q$ is $2$, all biprimes $n$ are odd and hence the least significant bit of $n,~p,\text{ and }q$ are $1$ i.e. $p_0 = q_0 = 1$. The most significant bits can also be set to $1$ by construction i.e. $p_{l_p-1}=q_{l_q-1}=1$.

\begin{table}
\begin{eqnarray}
\hspace*{-0.3cm}  \begin{array}{|r ||c|c| c|c|c|c|c|c|c|c|} 
\hline
& B_9 & B_8 & B_7 & B_6 & B_5 & B_4 & B_3 & B_2 & B_1 & B_0  \\ 
\hline
\hline
\multirow{2}{*}{\footnotesize }p=&  &   &    &    &    &   1  & p_3 & p_2 & p_1 & 1 \\ 
q=&  &  &     &    &   &   1  & q_3 & q_2 & q_1 & 1\\  
\hline
\hline
R_0~&  &  &    &   &   & 1  & p_3 & p_2 & p_1 & 1\\ \hline
R_1~&  &   &   &    &  q_1  & p_3q_1 & p_2q_1 & p_1q_1 & q_1 &   \\  \hline
R_2~&  &  &    &  q_2 & p_3q_2 & p_2q_2 & p_1q_2 & q_2 &     &   \\   \hline
R_3~&  &    &  q_3 & p_3q_3 & p_2q_3 & p_1q_3 & q_3 &     &   &\\   \hline
R_4~& & 1  & p_3 & p_2 & p_1 & 1 &   &     &  & \\
\hline
\multirow{2}{*} 
& _{8\rightarrow 9} & _{7\rightarrow 8} & _{6\rightarrow 7} & _{5\rightarrow 6} & _{4\rightarrow 5} & _{3\rightarrow 4} & _{2\rightarrow 3} &_{1\rightarrow 2} & & \\
\multirow{2}{*}{\footnotesize carry}& c_{89} & c_{78} & c_{67} & c_{56} & c_{45} & c_{34} & c_{23} & c_{12} & &  \\[4pt]

\multirow{2}{*} 
& _{7\rightarrow 9} & _{6\rightarrow 8} & _{5\rightarrow 7} & _{4\rightarrow 6} & _{3\rightarrow 5} & _{2\rightarrow 4} & & & & \\
& c_{79} & c_{68} & c_{57} & c_{46} & c_{35} & c_{24} & & & & \\[2pt]
\hline
\hline
\multirow{1}{*}{551}& 1 & 0 & 0 &  0 & 1& 0& 0& 1 & 1 & 1  \\
\hline
 \end{array} \nonumber
\end{eqnarray}
\caption{Bitwise multiplication table for $n =551=pq$ with $l_n = 10$, $l_p = l_q = 5$. $c_{ij}$ is the carry bit from column $i$ to column $j$.}
\label{tab1}
\end{table}
We set up the bitwise multiplication table and each column of the table gives rise to a factoring equation. An example for the said multiplication table is shown in Table I for the composite number $N= 551$ ($l_n=10)$ with factors $p=29$ ($l_p =5$) and $q=19$ ($l_q =5$) following the prescription in \cite{xu2012quantum}.
Here the first row indicates the bit-places and the subsequent two rows (having bit-variables $p_1$ to $p_3$ and $q_1$ to $q_3$) represent the two factors.  
The remaining rows indicate bit-wise products as well as the carry-bits ($c_{ij}$) from one column to another as indicated in the table. In the following we discuss how  factoring can be achieved using a hybrid computer with lesser number of qubits.

\subsection{Bitwise factoring equations}
It can be seen that there are two possible cases regarding the bitlengths of the factors, {\tt Case-A}: $l_n =l_p+l_q$ or {\tt Case-B}: $l_n = l_p+l_q-1$ .  Without loss of generality, assuming $q<p$, one can show that
$
l_q\leq\ceil{{l_n}/{2}}\leq l_p
$
 where $\ceil{\cdot}$ is the ceiling function. 
 Therefore, depending on the bit-size $l_n$ of the composite number, one may try various possibilities for the bit-sizes of factors, there can be at most $\ceil{l_n/2}$ of them. Typically, in cryptosystems which rely on the difficulty of prime-factorization, $l_p$ and $l_q$ are chosen to be comparable, else the factoring could be rendered easier. In the following we set up the factoring equations for general $l_p$ and $l_q$ and then eventually focus on the case where $l_p=l_q=\ceil{l_n/2}$.

First, it is important to note that not all the bits of the two factors contribute to $i$th bit of $n$.  
Since $n = pq$, 
\begin{eqnarray}
\sum_{i=0}^{l_n-1}2^i n_i =  \sum_{k=0}^{l_q-1}
\sum_{j=0}^{l_p-1}2^{j+k} p_j q_k.
\end{eqnarray}
Reshuffling the sum on the right hand side to collect terms with the same power of $2$, we have
\begin{eqnarray}
\sum_{i=0}^{l_n-1}2^i n_i = \sum_{m=0}^{l_p+l_q-2}2^{m} \sum_{k=\alpha_m}^{\beta_m}
 p_{m-k} q_k
 \label{reshuffled}
\end{eqnarray}
where $\alpha_m = \max(0,m-l_p+1)$ and $\beta_m  = \min(m,l_q-1)$.

At every order $m$ the sum $\sum p_{m-k} q_k$ can be broken up into a binary residue along with a carry variable (not necessarily binary) which adds to the terms in the next order $m+1$. By the same token, the m\textsuperscript{th} order will have an ``incoming'' carry variable $\f{m}$ from the $m-1$\textsuperscript{th} order. Thus the factoring stand as
	\begin{equation}
	\sum_{k=\alpha_m}^{\beta_m}
	 p_{m-k} q_k+ \f{m}=n_m+2\f{m+1}
	 \label{factoreq}
	\end{equation}
for $0\leq m\leq l_p+l_q-2$.
 The advantage is that unlike in the prescription in \cite{xu2012quantum} the factoring equations in \ref{factoreq} only couple adjacent orders, i.e. the $m^{\text{th}}$ equation gets connected only to the $m-1^{\text{th}}$  and $m+1^{\text{th}}$ equations. The trade-off is that these ``cumulative'' carry variables $\f{m}$ will in general take values in the set of non-negative integers. 

%
%
%
%
%
\begin{table}
\begin{eqnarray}
\hspace*{-0.3cm}  \begin{array}{|r ||c|c| c|c|c|c|c|c|c|c|} 
\hline
& B_9 & B_8 & B_7 & B_6 & B_5 & B_4 & B_3 & B_2 & B_1 & B_0  \\ 
\hline
\hline
p=&  &   &    &    &    &   1  & p_3 & p_2 & p_1 & 1 \\ 
q=&  &  &     &    &   &   1  & q_3 & q_2 & q_1 & 1\\  
\hline
\hline
R_0~&  &  &    &   &   & 1  & p_3 & p_2 & p_1 & 1\\ 
\hline 
R_1~&  &   &   &    &  q_1  & p_3q_1 & p_2q_1 & p_1q_1 & q_1 &   \\  \hline
R_2~&  &  &    &  q_2 & p_3q_2 & p_2q_2 & p_1q_2 & q_2 &     &   \\   \hline
R_3~&  &    &  q_3 & p_3q_3 & p_2q_3 & p_1q_3 & q_3 &     &   &\\   \hline
R_4~& & 1  & p_3 & p_2 & p_1 & 1 &   &     &  & \\
\hline 
\multirow{2}{*} 
& _{8\rightarrow 9} & _{7\rightarrow 8} & _{6\rightarrow 7} & _{5\rightarrow 6} & _{4\rightarrow 5} & _{3\rightarrow 4} & _{2\rightarrow 3} &_{1\rightarrow 2} & _{0\rightarrow 1} & \\
{\text{\footnotesize carry}}&  \f{9} & \f{8} & \f{7} & \f{6} & \f{5} & \f{4} & \f{3} & \f{2} & \f{1} & 0 \\[2pt]
\hline
\hline
\footnotesize 551 & 1 & 0 & 0 &  0 & 1& 0& 0& 1 & 1 & 1  \\
\hline
 \end{array} \nonumber
\end{eqnarray}
\caption{Bitwise multiplication table for $n =551=pq$. $\f{i}$ are the cumulative carries from column $i-1$ to column $i$.}
\label{tab2}
\end{table}
Note that $\f{0}\equiv0$ because the first column can't have an ``incoming'' carry. Furthermore, substituting \ref{factoreq} into \ref{reshuffled} we get
	\begin{equation*}
	\sum_{i=0}^{l_n-1}2^i n_i = \sum_{m=0}^{l_p+l_q-2}2^{m} n_m + 2^{l_p+l_q-1}\f{l_p+l_q-1}
	\end{equation*}
from which we can conclude that 
	\begin{equation}
	\f{l_p+l_q-1}= %
	\begin{cases}
		n_{l_n-1}=1& \text{for {\tt Case-A}: } l_n =l_p+l_q\\
		0    & \text{for {\tt Case-B}: } l_n =l_p+l_q-1
	\end{cases} \nonumber
	\end{equation}
From the structure of the factoring equations it is possible to readily assign values to some of the $\f{i}$, namely $\f{1}$ and $\f{l_p+l_q-2}$
	\begin{align*}
	m&=0\hphantom{l_p+l_q-2}:
	1=1+2\f{1}\Rightarrow \f{1}=0\\
	m&=l_p+l_q-2\hphantom{0}:
	\f{l_p+l_q-2}=n_{l_p+l_q-2}+2\f{l_p+l_q-1}-1
	\end{align*}

The factoring equations can be put into a convenient matrix form as well. For concreteness, for $n=551$, the matrix representation of Eq. \ref{factoreq} is    
\begin{eqnarray}
     \begin{bmatrix}
       1 & 0 & 0 & 0 & 0\\
       q_1 & 1 & 0 & 0 & 0\\
       q_2 & q_1 & 1 & 0 & 0\\
       q_3 & q_2 & q_1 & 1 & 0\\
       1 & q_3 & q_2 & q_1 & 1 \\
       0 & 1 & q_3 & q_2 & q_1 \\
       0 & 0 & 1 & q_3 & q_2\\
       0 & 0 & 0 & 1 & q_3\\
       0 & 0 &0 & 0 & 1\\
       0 & 0 &0 & 0 & 0\\
       \end{bmatrix}
       \begin{bmatrix}
       1\\
       p_1\\
       p_2\\
       p_3\\
       1\\
       \end{bmatrix}
       +
       \begin{bmatrix}
       0\\
       0\\
       \f{2}\\
       \f{3}\\
       \f{4}\\
       \f{5}\\
       \f{6}\\
       \f{7}\\
       \f{8}\\
       \f{9}\\
       \end{bmatrix}
      =
       \begin{bmatrix}
       1\\
       1\\
       1\\
       0\\
       0\\
       1\\
       0\\
       0\\
       0\\
       1\\
       \end{bmatrix}
       +
       2 
       \begin{bmatrix}
          0\\
          \f{2}\\
          \f{3}\\
          \f{4}\\
                 \f{5}\\
                 \f{6}\\
                 \f{7}\\
                 \f{8}\\
                 \f{9}\\
          0\\
       \end{bmatrix}.
       ~~~  
       \label{fe}
\end{eqnarray}
Thus a general factoring problem can be converted into solving equations of the above structure.   

\subsection{\label{sec:level2} Simplifying bitwise factoring equations via classical processor}
Even though the $\f{i}$ variables aren't binary, it is possible to place bounds on them by noting that 
\begin{eqnarray}
\max[\f{i+1}] = \floor{\frac{1}{2}\max \left( \sum_{k = \alpha_i}^{\beta_i}p_{i-k}q_k + \f{i} \right)- \frac{n_i}{2}}
\label{maxCi}
\end{eqnarray}
,where $\floor{\cdot}$ denotes the floor function. This is arrived at from rearranging the factoring equations. It is also possible to inductively determine an absolute upper bound for individual $\f{i}$ irrespective of $n_i$, namely,
\begin{equation}
\max[\f{i}]	= 
\begin{cases}
i-1& \text{for }1\leq i\leq l_q-1 \\
l_q-1& \text{for }l_q\leq i\leq l_p \\
l_p+l_q-i& \text{for }l_p+1\leq i\leq l_p+l_q-2.
\end{cases} 
\label{maxc}
\end{equation}
The above values are used to initialize the $\{\f{i}\}$ and then the bound on each element can be iteratively refined using eq. \ref{maxCi}, where the maximum over the binary variables $\{p_i,q_i\}$ are evaluated in accordance with the constraints between them.

In the case of $n=551$, considering column $B_1$ from Table \ref{tab2}
we find that $p_1 + q_1 = 1 + 2\f{2}$ while,
\begin{eqnarray}
\max[\f{2}] = \floor{\frac{1}{2} \left(\max\sum_{j = 0}^{1}p_{1-j}q_j + \max[\f{1}]-n_1\right)}= 0 \nonumber,
\end{eqnarray}
since $\alpha_1 = \max(0,  1-10+5+1) = 0$, $\beta_1 = \min(1,4) = 1$, and
therefore $\f2=0$. In the same way, using bitwise logic, the classical processor can determine values of all other $\{\f{i}\}$s.  For $n=551$, using a simple numerical procedure we found that
\begin{eqnarray}
 &&\f{3} = 0, ~\f{4} = 1, ~ \f{5} = 2, ~ \f{6} = 1, \nonumber \\
 && \f{7} = 1, ~ \f{8} = 1, \mbox{ and } \f{9} = 1.
 \label{f}
 \end{eqnarray}
 The simplified matrix representation of the relevant factoring equations now becomes,
\begin{eqnarray}
\begin{bmatrix}
q_1 & 1 & 0 & 0 \\
q_2 & 0 & 1 & 0 \\
q_3 & 0 & 0 & 1\\
0 & q_2 & q_1 & 0 \\
0 & q_3 & 0 & q_1 \\
0 & 0 & q_3 & q_2 \\
\end{bmatrix}
\begin{bmatrix}
1\\
p_1\\
p_2\\
p_3\\
\end{bmatrix}
=
\begin{bmatrix}
1\\
1\\
1\\
1\\
1\\
0\\
\end{bmatrix}.
\end{eqnarray}
Since they involve six unknowns, namely $\{p_1,p_2,p_3\}$ and $\{q_1,q_2,q_3\}$, it takes six variables to factor 551.  However, a further reduction in number of variables is possible by exploiting the first three equations, namely $p_1+q_1 = 1$, $p_2+q_2 = 1$, and $p_3+q_3 = 1$, which together imply that $q_j = 1-p_j$.  Finally only three unknowns define the factoring equations:
\begin{eqnarray}
p_1 (1-p_2) + (1-p_1) p_2-1 &=& 0, \nonumber \\ 
p_1 (1-p_3) + (1-p_1) p_3 - 1 &=& 0, ~\mathrm{and} \nonumber \\
(1-p_2) p_3  +  p_2(1-p_3) &=& 0.
\label{faceq}
 \end{eqnarray}
In the following we describe how these equations are solved using a 3-qubit adiabatic quantum processor.

\subsection{\label{sec:level2}Solving the bitwise factoring equations via quantum adiabatic processor}
\subsubsection{Quantum Adiabatic Algorithm}
Consider a closed quantum system existing  in an eigenstate $\ket{\psi_i}$ of the initial Hamiltonian ${\mathcal H}_i$ which is slowly changed to a new Hamiltonian ${\mathcal H}_f$.
Then, according to the quantum adiabatic theorem, the system mostly remains in an eigenstate of the instantaneous Hamiltonian  and ultimately reaches the corresponding eigenstate of the final Hamiltonian, provided the system does not find two or more crossing eigenstates during the process \cite{nielsen2010quantum, farhi}.

Given a problem, adiabatic quantum computation typically involves encoding the solution to the problem in the ground-state of the final Hamiltonian.  A suitable initial 
Hamiltonian is chosen for which ground-state can be prepared easily.   Then the Hamiltonian of the system is slowly varied such that the system stays in the ground state of the instantaneous Hamiltonian. 
The intermediate Hamiltonian can be seen as an interpolation (linear or nonlinear)  between the initial and final Hamiltonian \cite{hu2016optimizing}.  If $T$ is the total time of evolution and $0 \le s \le 1$ is the interpolation parameter, then
\begin{eqnarray}
{\mathcal H}(s) = (1-s){\mathcal H}_i + s{\mathcal H}_f.
\end{eqnarray} 
For linear interpolation we choose $s=t/T$, where $t$ is the instantaneous time of evolution \cite{messiah1958quantum}.
The Adiabatic theorem requires that
\begin{eqnarray}
T = \left\vert \frac{\max \{d {\mathcal H}(s) /ds\}}{\epsilon\Delta^2/\hbar} \right\vert,
\end{eqnarray}
where $\Delta$ is the minimum energy gap between the ground and the first excited state.  Probability of reaching the ground state of the final Hamiltonian is given by $1 - \epsilon^2$. From here onwards, we set $\hbar=1$ and express Hamiltonian in angular frequency units. 

Now the entire time evolution of the  system from ${\mathcal H}_i$ to ${\mathcal H}_f$ can be thought of a unitary transformation $U_T$ generated by a piece-wise-constant Hamiltonian 
\begin{eqnarray}
{\mathcal H}_m = (1 - m/M){\mathcal H}_i+(m/M){\mathcal H}_f,
\end{eqnarray}
with $M$ pieces, each of duration $\tau$, and $0 \le m \le M$.
Defining $U_m = \exp(-i {\mathcal H}_m \tau)$, the total evolution operator $U_T = \prod_{m=1}^{M}U_m$.  

\subsubsection{Quantum Adiabatic Factoring}
In order to convert the factorization problem into an optimization problem, Peng et al constructed a cost function
$f(p,q) = (n-p \cdot q)^2$ which is minimum when $p$ and $q$ are the factors \cite{pengfactorizationAQC}.  They replace the scalar variables $p$ and $q$ with operators 
\begin{eqnarray}
P = \sum_{i=0}^{l_p-1}2^i W_i
~~\mathrm{and} ~~
Q = \sum_{i=0}^{l_q-1}2^i W_i.
\end{eqnarray}
Here the number operator $W_i = (I_2-\sigma_{iz})/2$ is constructed in terms of the identity operator $I_2$ and the 
Pauli z-operator $\sigma_z$ of the $i$th qubit.
Note that eigenvectors $\ket{0}$ and $\ket{1}$ of $W_i$ have the eigenvalues $0$ and $1$, the values a classical bit can take. 
Using this method,
Peng et all could factor the number 21 
from the adiabatically prepared ground state of the final Hamiltonian
\begin{eqnarray}
{\cal H}_f = (N I_{2^n} - P \cdot Q)^2.
\label{pengHf}
\end{eqnarray}
It can be noted that the ground state of the above Hamiltonian represents the factors.
However, extending this method for factorizing larger numbers is difficult since the Hamiltonian in Eq. \ref{pengHf} can have many-body terms and required a large number of qubits.

Du et al improved upon this scheme using Table I \cite{xu2012quantum}. Each column of Table I represents an equation which is subsequently encoded into a bit-wise Hamiltonian, whose ground state contains the information about  respective bits of the two factors.  For example, 
\begin{eqnarray}
&&B_1:~~ p_1+q_1-1-2c_{12} = 0, \nonumber \\
&&B_2:~~ p_2+p_1q_1+q_2+c
_{12}-1-2c_{23}-4c_{24}=0, \nonumber
\end{eqnarray}
and so on.

Now the bit variables are replaced by the number operators:  $p_j \rightarrow W_j$, $q_j \rightarrow W_{j+l_p-2}$. The carry bits $\{c_{j,j+i}\}$ are organised in a list according to increasing $i$ for the same $j$ and then in order of increasing $j$. Each element $k$ of the list is mapped onto $W_{k+l_q+l_p-4}$.   
The bit-wise Hamiltonians are then
\begin{eqnarray}
&&B_1:~ {\cal H}_1 = (W_1 + W_4 - 1 - 2W_7)^2, \nonumber \\
&&B_2:~ {\cal H}_2 = (W_2+W_1W_4+W_5+W_7-1-2W_8-4W_{15})^2, \nonumber
\end{eqnarray}
and so on.  
Thus, the final Hamiltonian of the factorization problem is the sum
\begin{eqnarray}
{\cal H}_f = \sum_{i=1}^{l_n-1} {\cal H}_i.
\label{Hf}
\end{eqnarray}
Some of these terms ${\cal H}_i$ will involve 4-body interactions which can be reduced to 3-body interactions following the prescription outlined in \cite{schaller2007role}. If the Hamiltonian is varied slowly enough, the adiabatic theorem ensures that the system ends up, with high probability, in the ground state of the target Hamiltonian. 
Therefore on measuring the adiabatically prepared ground state of ${\cal H}_f$, it is possible to retrieve the factors.  Although the above encoding requires 20 qubits to factor the number 551, our hybrid scheme (section II B) requires only 3 qubits.  

\subsubsection{Quantum Adiabatic Factoring of 551}
In a hybrid computer, we first reduce the bitwise factoring equations as described in section II B and then apply the quantum adiabatic algorithm to solve the residual equations.  For the specific case of 551, the factoring equations are given by Eq. \ref{faceq}. Replacing $p_j \rightarrow W_j = (I_2-\sigma_{jz})/2$, we form the bitwise Hamiltonian ${\cal H}_i$.  Final Hamiltonian (Eq. \ref{Hf}) becomes,
              \begin{eqnarray}
              H_f = (3I_8+ \sigma_z^1\sigma_z^2 - \sigma_z^2\sigma_z^3 + \sigma_z^1\sigma_z^3)/2.          \end{eqnarray}
              
In the next section we describe the experimental determination of the ground state of the above Hamiltonian which reveals the factors of 551.            

%
%
       
\section{\label{sec:level3}Experiment}
We implement the adiabatic factorization of 551 on a three qubit NMR register involving $^1$H, $^{19}$F, and $^{13}$C of dibromofluoromethane (DBFM)  dissolved in acetone-D6 \cite{mitra2008nmr}. All the experiments were carried out on a Bruker 500 MHz NMR spectrometer at an ambient temperature of 300 K. 

\begin{figure}
	\centering
	\includegraphics[trim=0.3cm 0cm 0cm 0cm, clip=true,width=8.5cm]{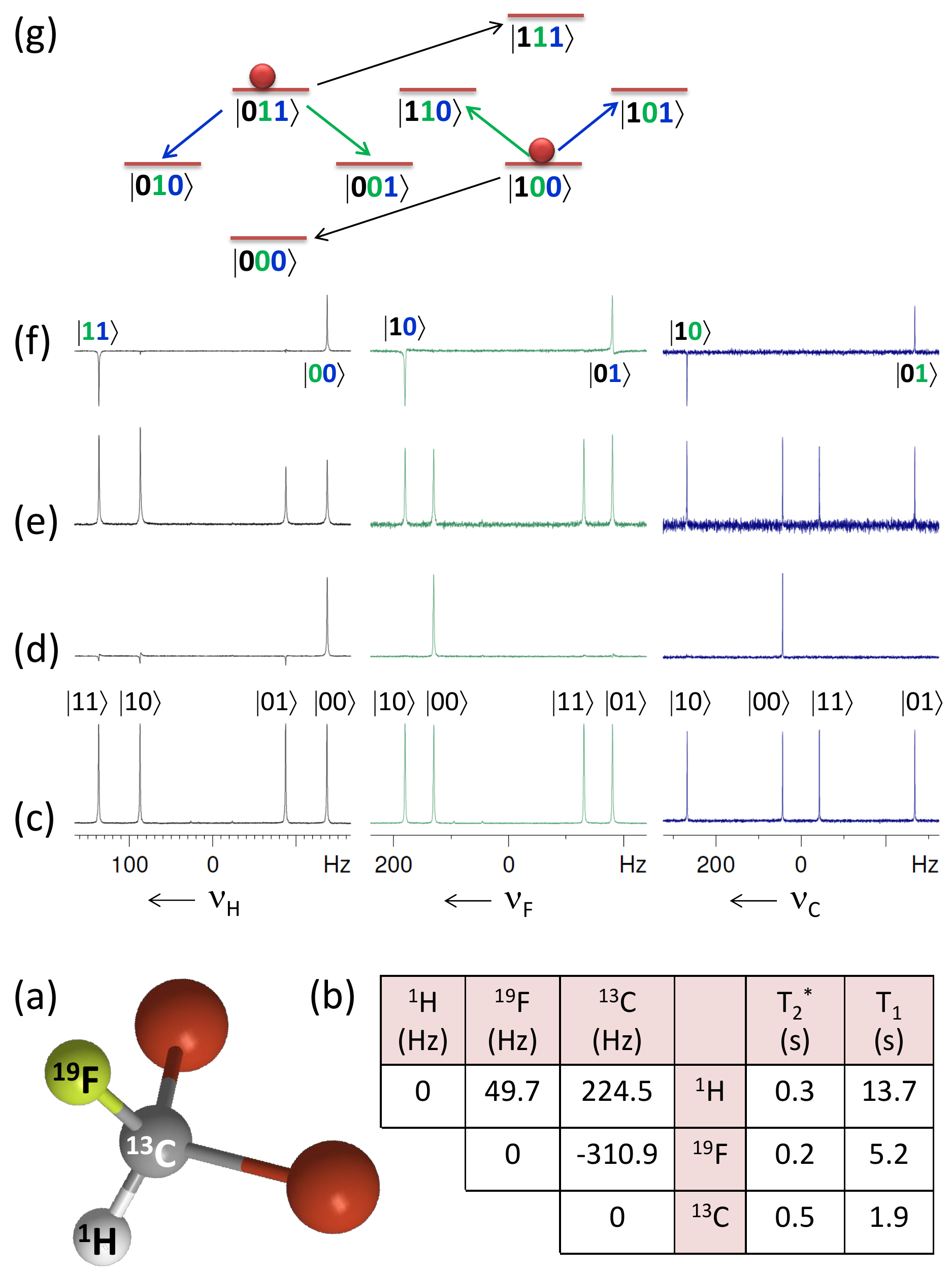}
	\caption{The molecular structure of dibromofluoromethane is shown in (a). Resonance offsets ($\nu_i$; diagonal elements), coupling constants ($J_{ij}$; off-diagonal elements) and relaxation parameters are tabulated in (b). The experimental NMR spectra correspond to thermal equilibrium (c), PPS (d), the ground state of initial Hamiltonian ${\cal H}_i$ (e), and the solution, i.e., the ground state of the final Hamiltonian  ${\cal H}_f$ (f).  The energy-level diagram (g) describes the deviation populations in the final state.}
	\label{dbfm}
\end{figure}

The internal Hamiltonian for the three-qubit system under week-coupling approximation \cite{cavanagh,levittbook}, can be written as 
\begin{eqnarray}
{\cal H}_\mathrm{int} = -2\pi \sum_{i=1}^{3}\nu_i I_z^i + 2\pi \sum_{i=1,j>i}^{i=2}J_{ij}I_z^iI_z^j
\end{eqnarray}
Where $\nu_i$ are the resonance offsets, $J_{ij}$ are the coupling constants, and $I_z^i$ are the z-components of spin angular momentum operators. 
The molecular structure, Hamiltonian parameters and the thermal equilibrium  spectra of DBFM are shown in Fig. \ref{dbfm}(a-c) respectively. 
 
\begin{figure}
	\centering
	\includegraphics[trim=3cm 0cm 4cm 0.5cm,clip=true,width=8cm]{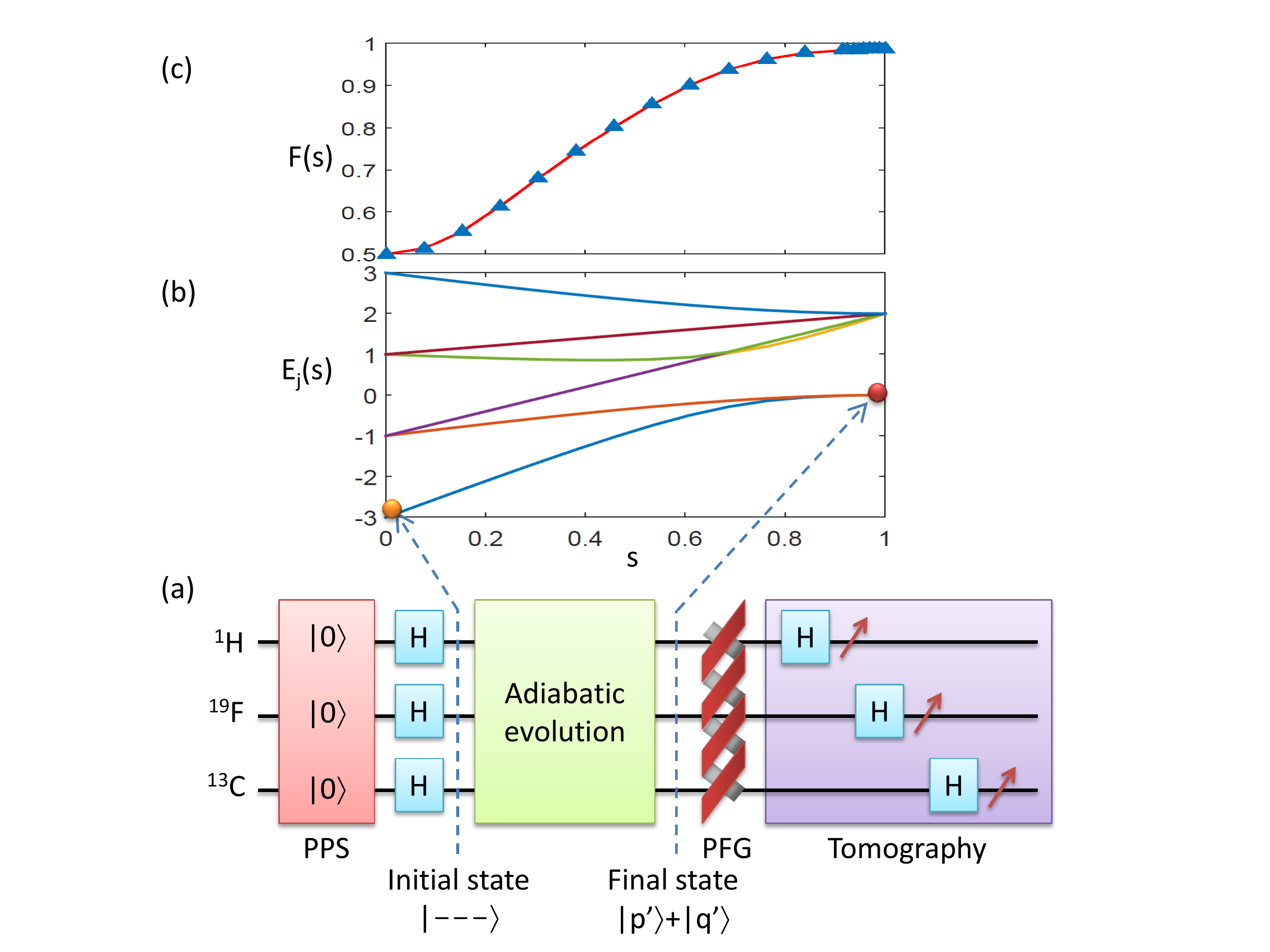}
	\caption{Three qubit circuit for solving the bitwise factoring equations (a), the transformation of energy spectrum during the adiabatic evolution (b), and the simulated fidelity of the solution state with the instantaneous ground state during the adiabatic evolution (c).}
	\label{circuit}
\end{figure} 
 
The complete circuit for the experiment is shown in Fig. \ref{circuit}. The experiment mainly involves the following four stages.
\begin{itemize}
\item[(i)]{Initialization:
Preparation of $\ket{000}$ pseudopure state (PPS) from thermal equilibrium state was achieved by standard methods \cite{corypps,chuangpps,PhysRevLett.106.080401}.  The PPS spectra shown in Fig. \ref{dbfm}d corresponds to a fidelity of over 0.99.}
\item[(ii)]{Preparing the ground state: We choose the initial 
Hamiltonian to be 
\begin{eqnarray}
{\cal H}_i = \sigma_{x}^1+\sigma_{x}^2+\sigma_{x}^3, \end{eqnarray}
whose ground state is $\ket{---}$ (where $\ket{\pm} = (\ket{0}\pm \ket{1})/\sqrt{2}$).
Transforming the PPS into $\ket{---}$ was achieved by using three pseudo-Hadamard gates ($H = \exp[i(\pi/2)\sigma_y/2]$)
and the corresponding experimental spectra are shown in Fig. \ref{dbfm}e.}
\item[(iii)]{Adiabatic evolution: The ground state of the initial Hamiltonian was driven adiabatically towards the ground state of the final Hamiltonian  ${\cal H}_f$ (as in Eq. \ref{Hf}) over a duration $T=3.5$ s in 20 steps.  The progression of energy eigenvalues $E_j(s)$ as a function of the interpolation parameter $s$ is shown in Fig. \ref{circuit}(b).  Note that the ground state has no cross-over except at the end of the evolution where it becomes doubly degenerate.  Each of these degenerate eigenstates encodes a factor.  To quantify the overlap between 
the expected probabilities $p^\mathrm{th}_j$ and the simulated probabilities $p^s_j$ after the $s$th step
we define a fidelity measure
\begin{eqnarray}
F(s) = \frac{\sum_j p^\mathrm{th}_j p^s_j}{\sqrt{\sum_j (p^\mathrm{th}_j)^2 \sum_j (p^s_j})^2 }.
\end{eqnarray}
The profile of $F(s)$ versus the interpolation parameter $s$ ultimately reaches a value of 0.99 at the end of evolution (see Fig. \ref{circuit}(c)).    

The propagators corresponding to these adiabatic steps are realized using the recently developed Bang-Bang quantum control technique \cite{GauravBB}. The obtained RF sequences were robust within an RF inhomogeneity of $\pm 10 \%$ and had average fidelities above 0.99.  
	}
	
\begin{figure}
	\centering
	\includegraphics[trim=4cm 0cm 5cm 0cm, clip=true,width=8cm]{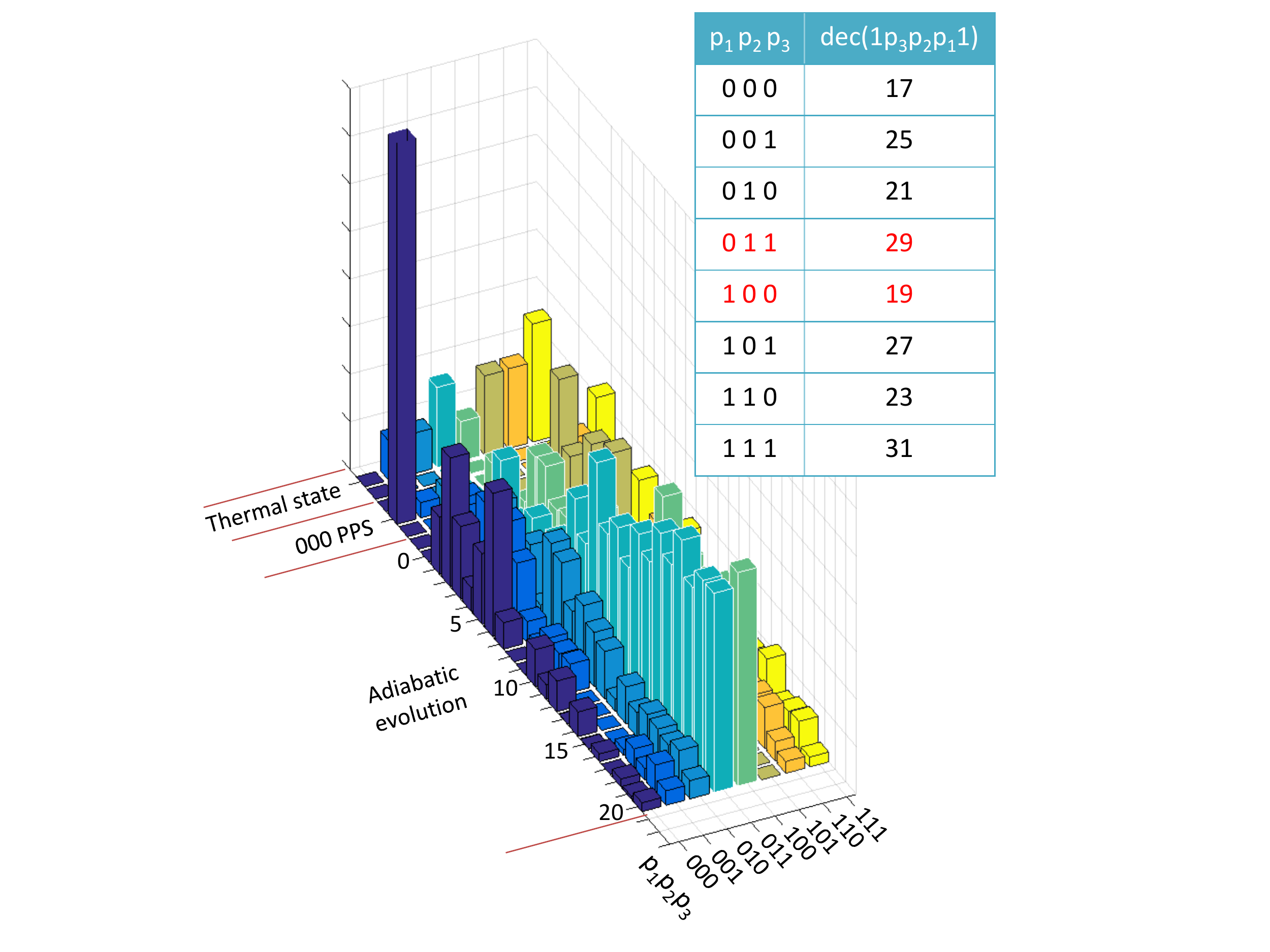} 
	\caption{Experimental probabilities of all the eight eigenstates at various stages of circuit in Fig. 2.  Evolution of the probabilities during all the 20 adiabatic steps is shown.  The table describes decoding the various eigenstates into respective decimal numbers.  Factors highlighted in red achieve the highest probabilities during the adiabatic process.}  
	\label{tomo1}
\end{figure}

\item[(iv)]{Measurement of probabilities: 
    To demonstrate the evolution of the probabilities during the adiabatic process, we carried out 20 experiments each with varying length of the adiabatic sequence.  In each experiment, after dephasing the coherences using a pulsed-field gradient (PFG) \cite{jones2009magnetic}, we measured the probabilities of various eigenstates in the computational basis (see Fig. \ref{circuit} (a)) \cite{chuangtomo,maheshtomo}. The barplots of the probabilities versus the number of steps is shown in Fig. \ref{tomo1}.
    
    The experimental spectra of the final state and the corresponding population distributions are shown in Fig. \ref{dbfm}(f) and 1(g) respectively.  The fidelity of final state with the desired target state was over 0.99.}
\end{itemize}

\textit{Discussions:}
It is clear from the table in Fig. \ref{tomo1} that the final state encodes the factors 19 and 29 with high probability.  As with an NP problem, these factors can be verified easily.  

An important issue is the complexity of the whole process, which is discussed qualitatively in the following.  Formulating the bitwise factoring equations (Eq. \ref{fe}) involve  mainly bit-wise multiplications, and hence  polynomial in the bit-size of the composite number ($l_n$).  In principle, these factoring equations can directly be passed on to a quantum processor with a large number of qubits. Instead, we used some simple classical routines to reduce the size of the quantum register.  This procedure involves computing upper bounds of cumulative carries $C_i$ (see Eqs. \ref{maxc}) and its complexity depends on the particular classical algorithm used.  We presume that this optional procedure can be carried out efficiently without any exponential complexity.  The quantum adiabatic process for solving the linear equations itself is believed to be polynomial \cite{FarhiAQC,AQC}.  Therefore, we believe that the overall factorization procedure is efficient.

The crucial point in a hybrid scheme is to maximize the efficiency of the overall computation by optimizing the switching point from classical to quantum processor.
 In this particular problem, simplifying the factoring equations to a higher extent will mean lesser number of required qubits during the quantum procedure. However, the complexity of classical simplification by itself should remain polynomial. The exact point of crossover depends on the particular problem at hand and needs further investigation.
 
In the case factoring $551$, it so happened that calculating upper and lower bounds of carries $\f{i}$ were enough to fix the values of the same. However, it is probable that for larger numbers, this procedure may not be able to fix  the values of all the carry variables, and the simplified factoring equations which are passed to the quantum routine may involve those unknown carries $\f{i}$. Nevertheless, these variables will be bounded from above and below, making the number of qubits required to encode them less than in the unbounded case. 

\section{Conclusions}
Although, classical computers have seen an enormous progress over the past few decades, their difficulty in factorization has become the corner stone of classical cryptography.  Quantum computers are capable of factoring large numbers with polynomial complexity.  Although prototype quantum computers capable of factoring small numbers have already been built, a large quantum computer outperforming a classical computer is just not around the corner.   In this scenario, it is possibly more realistic to look for a hybrid computer having both classical and quantum processors.  

In the present work, we analyzed a possible scheme to factor a composite number by combining certain bitwise operations using a classical processor, and then solving a set of linear equations using an adiabatic quantum processor.  We described the algorithm with respect to factoring the number 551 into 19 and 29 using only three qubits.  Finally we experimentally demonstrated the adiabatic quantum algorithm using a three-qubit NMR quantum simulator, and obtained the factors with high probability.  We believe this as a first step in exploiting the best of both the classical and quantum computational capabilities.

\section*{Acknowledgement}
Authors acknowledge useful discussions with Sudheer Kumar and Abhishek Shukla.   SP and SM acknowledge hospitality from Indian Institute of Science where this work was initiated. SM would like to thank Indian Academy of Sciences for the support during this period. 
This work is supported by Department of Science and Technology, India (grant number DST/SJF/PSA-03/2012-13) and Council of Scientific and Industrial Research, India (grant number CSIR-03(1345)/16/EMR-II).  


\bibliographystyle{apsrev4-1}
\bibliography{draft6}
\end{document}